\documentstyle[preprint,revtex]{aps}
\def\L{\Lambda}
\def\l{\lambda}
\def\Ls{\tilde\Lambda}   % : Lambda sombrero
\def\ls{\hat\lambda}
\def\ket{\rangle}
\def\ol{\overline}
\def\beq{\begin{equation}}
\def\eeq{\end{equation}}
\def\beqa{\begin{eqnarray}}
\def\eeqa{\end{eqnarray}}
\def\G{SU(m+1)}
\def\H{SU(m)\times U(1)}
\def\SUM{SU(m)}
\def\c{\chi_}
\def\D{\Delta}
\def\s{\sigma}
\def\w{\omega}
\def\N{\cal N}
\def\M{\cal M}
\def\P{{\cal{P}}}
\def\nn{\nonumber}
\def\bc{|cr|^2}
\def\t{\bar t}

\begin{document}

\begin{title}
NONDIAGONAL $CP_m$  COSET MODELS \\
AND THEIR POINCAR\'E  POLYNOMIALS
\end{title}

\author{G. Aldazabal, I. Allekotte}
\begin{instit}
Centro At\'omico Bariloche$^*$,
8400 S. C. de Bariloche,\\
Comisi\'on Nacional de Energ\'{\i}a At\'omica,\\
Consejo Nacional de Investigaciones Cient\'{\i}ficas
y T\'ecnicas,\\
 and Instituto Balseiro, Universidad Nacional de Cuyo, Argentina.
\end{instit}

\author{E. Andr\'es}
\begin{instit}
Centro At\'omico Bariloche,
8400 S. C. de Bariloche,  \\
Comisi\'on Nacional de Energ\'{\i}a At\'omica,\\
 and Instituto Balseiro, Universidad Nacional de Cuyo, Argentina.
\end{instit}

\author{C. N\'u\~nez}
\begin{instit}
Instituto de Astronom\'{\i}a y F\'{\i}sica del Espacio,
C.C. 67, suc. 28, 1428 Buenos Aires, \\
and Consejo Nacional de Investigaciones Cient\'{\i}ficas
y T\'ecnicas, Argentina.
\end{instit}

\begin{abstract}
 $N=2$ coset models of the type $SU(m+1)/SU(m)\times U(1)$ with nondiagonal
modular invariants for both  $SU(m+1)$ and $SU(m)$ are considered.
Poincar\'e polynomials of the corresponding chiral rings of these
algebras are constructed. They are used to compute the number of chiral
generations of the associated string compactifications. Moddings
by discrete symmetries are also discussed.
\end{abstract}

%%%%%%%%%%%%%%%%%%%%%%%%
%\section{Introduction} %
%%%%%%%%%%%%%%%%%%%%%%%%

\noindent{\bf 1. Introduction}

Critical superstrings in four spacetime dimensions may  be obtained by
different compactification procedures from the ten dimensional superstring.
$N=1$ spacetime supersymmetry in four dimensions requires a conformal field
theory with N=2 supersymmetry in the internal compactified
sector~$^{\cite{fksw}}$. An interesting alternative way of
achieving consistent string theories in four dimensions was suggested by
D. Gepner~$^{\cite{gepner}}$. In this construction, known as ``$N=2$ string
theory,"  the internal sector corresponds to a tensor product of $N=2$
superconformal minimal models with total central charge $c_{int}= 9$.
The spacetime supersymmetry projector is constructed from the $U(1)$ current
of the algebra. Compatibility with modular invariance is ensured by projection
over states with odd integer charge. In order to obtain a gauge group
large enough to include the standard model, a heterotic transformation
on the left four dimensional spacetime sector can be carried out without
spoiling modular invariance. In a generalization of this scheme, provided
by Kazama and Suzuki in Ref.~\cite{ks}, the internal sector is built out of
tensor products of coset models.

$N=2$ string theory provides an algebraic construction of four dimensional
strings. Information of this algebraic structure can be encoded in a
Poincar\'e polynomial which counts chiral primary states together with
their $U(1)$ charge. The construction of these Poincar\'e polynomials
is relevant for finding the relation between this algebraic approach
and the geometric Calabi-Yau compactifications. This relation has been
established in some cases in Refs.~\cite{gepsft,gvw,lvw}.
In this article we compute Poincar\'e polynomials for $CP_m$ coset models,
{\sl i.e.}  $SU(m+1)/SU(m)\times U(1)$ models, coupling the left and right
moving sectors with nondiagonal invariants both for $SU(m+1)$ and $SU(m)$.

Due to the enormous degeneracy of four-dimensional string vacua
and the lack of a theoretical argument to pinpoint the correct one,
much effort~$^{\cite{models,butu,schell,nos,modd}}$ has been invested in a
systematic study of possible $N=2$ string models. In Ref.~\cite{nos}
the number of $E_6$ generations for $CP_m$ coset models
with a large class of nondiagonal couplings for the left and right
moving sectors of $SU(m+1)$ was computed by direct construction of
the massless spectrum.

The knowledge of Poincar\'e polynomials is particularly useful for
performing this kind of computations as shown by C. Vafa in Ref.~\cite{vafa}
and applied to this context by E. Buturovic $^{\cite{butu}}$.
Following these authors we further generalize the results of Ref.~\cite{nos}
by computing the number of chiral generations of $(2,2)$ string
compactifications with $CP_m$ coset models which have at least one
nondiagonal invariant for $SU(m)$ as well.

The number of $E_6$ generations obtained in this way
is too large to make these
models phenomenologically tractable. When moddings by discrete symmetries
are taken into account, this number is considerably reduced.
We therefore conclude by discussing moddings and describing
them in terms of Poincar\'e polynomials for the twisted sectors.

 ~~
%%%%%%%%%%%%%%%%%%%%%%%%%%%%%%%%%%%%%%
%\section{$N=2$ $CP_m$ Coset Models}  %
%%%%%%%%%%%%%%%%%%%%%%%%%%%%%%%%%%%%%%

\noindent{\bf 2.} ${\bf N=2}$ ${\bf CP_m}$ {\bf Coset Models}

Let us consider the quotient theory
$SU(m+1)_k\times SO(2m)_1 /SU(m)_{k+1}\times U(1)$ to which we shall
refer as $(m,k)$. We denote the fundamental weight vectors of $\G$ and
$\SUM$ by $\w_i$ with $i$ ranging from $0$ to $m$ and $m-1$ respectively.
States of the $N=2$ left superconformal algebra (SCA) are labelled by
$|\L,\l,\Ls\ket$, where $\L$ is a weight vector of $\G$ at level $k$
($\L=\sum_{i=1}^{m} m_i \w_i$; $0\leq \sum_{i=1}^{m} m_i\leq k$);
 $\Ls$ is a $SO(2m)$ weight at level 1 (so it can only take the values
$0$, $v$, $s$, $\ol s$) and $\l$ is a weight vector of $\H$ at level $k+1$
obtained by decomposing $|\L\ket\otimes|\Ls\ket$ into irreducible
representations of $\H$. Furthermore, $\l$ will be decomposed into a
$\SUM$ weight $\ls=\sum_{i=1}^{m-1} n_i \w_i$ and a $U(1)$ charge $q$
(corresponding to the $U(1)$ of $\H$) as:
\beq
\l=\ls +{\w_m\over m}\,q
\eeq

The conformal dimension $\D$ and $U(1)$ supersymmetry charge $Q$ are given by
\beqa \label{dim}
\D&=& {\L (\L + 2 \rho_{m+1})- \l(\l+2 \rho_m) \over 2 (k+m+1)}+
{\Ls^2\over 2} + N\\
\label{carga}
Q&=& -\sum_{l=1}^{2m} \Ls^l +
{2\over k+m+1}(\rho_{m+1}-\rho_m).\l+2M=\nonumber \\
 ~&=& -\sum_{l=1}^{2m} \Ls^l + {q\over k+m+1} +2M
\eeqa
where $\rho_{m+1}$ and $\rho_m$ denote half the sum of the positive
roots of $\G$ and $\SUM$ respectively,
and $N,M$ are integers.

However, not all the states constructed in this way are independent. This is
due to a state identification implied by the existence of a proper
external automorphism $\s$ of the extended Dynkin diagram of the
affine Lie algebra of $\G$.
Under this automorphism a state $|\L,\ls,q,\Ls\ket$ changes to
$^{\cite{gepnerfi,sierra,caihe}}$
$|\s(\L),\s(\ls),\s(q),\s(\Ls)\ket$, where
\beqa
\s(\L)  &=&  (k-\sum_{i=1}^{m} n_i)\w_1 + \sum_{i=2}^{m} n_{i-1} \w_i\nn\\
\s(\ls)  &=&  (k-\sum_{i=1}^{m-1} n_i)\w_1 + \sum_{i=2}^{m-1} n_{i-1} \w_i\nn\\
\s(q)   &=&  q+k+m+1 \nn\\
\s(\Ls=(0),(v),(s),(\bar s)) &=& ((v),(0),(\bar s), (s))
\label{fi}\eeqa
These states should be identified for the fields in the theory
to form a unitary representation of the modular group.
As the $CP_m$ models do not possess automorphism fixed points,
this field identification poses no further problem (for a discussion
of fixed points in coset models see Refs.~\cite{schell,fpp}).

Characters of the $N=2$ SCA are defined by
\beq
\chi^G_{\L} \, \chi^{SO(d)}_{\Ls}
=\sum_{\l} \, \chi^{N=2}_{\L,\l,\Ls} \, \chi^H_{\l}
\eeq

Under modular transformations
\beqa
\chi^{N=2}_{\L,\l,\Ls}({-1/\tau})= S_{\L,\L'} \, S_{\Ls,\Ls'}\,
S^*_{\l,\l'}\, \chi^{N=2}_{\L,\l,\Ls}(\tau) \nonumber \\
\chi^{N=2}_{\L,\l,\Ls}(\tau+1)= T_{\L,\L'}\,T_{\Ls,\Ls'}\,
T^*_{\l,\l'}\, \chi^{N=2}_{\L,\l,\Ls}(\tau)
\eeqa

Thus, a general modular invariant partition function for a given coset model
has the form
\beq
Z=\sum_{\L,\l,\bar\L,\bar\l,\Ls}\, \chi^{N=2}_{\L,\l,\Ls}\,\,
{\cal N}_{\L,\bar\L}\,\,
{\cal M}_{\l,\bar\l}\,\,\chi^{N=2\;*}_{\bar\L,\bar\l,\Ls}
\eeq
In the above equation the sum extends over states
$(\L,\l)$ and $(\bar\L,\bar\l)$
satisfying the condition $C(\L-\l)$ of Ref.~\cite{gepnerfi}, which for the NS
sector corresponds to $\L-\l \in M$, the root lattice of $\G$.
{$\cal N$} and {$\cal M$} denote modular invariants for
$\G$ and $\H$ respectively.
Up to date, a complete classification of modular invariants
for $SU(N)$, $N>2$, is still lacking. Nevertheless, in Ref.~\cite{nos}
a wide variety of them was used in model building. Only one invariant,
which is relevant to string compactification
when nondiagonal invariants for $\SUM$ are called for,
was not explicitly calculated before.
Namely, the invariant $C(5,8)$ derived from the embedding
of $SU(6)_8$ into $SO(21)_1$ $^{\cite{embedding}}$, which we list
in table I. For the remaining invariants we refer the reader to
Ref.~\cite{nos}.
For all invariants considered here the condition
${\N}_{\L,\L'}={\N}_{\s(\L),\s(\L')}$ is fulfilled.
In the above reference, the $SU(4)_{k odd}$ $G$ invariant does not satisfy
this relation. However, in this case all results are identical
to those obtained with the diagonal one.

  ~~

%%%%%%%%%%%%%%%%%%%%%%%%%%%%%%%%%%%
%\section{Poincar\'e Polynomials}  %
%%%%%%%%%%%%%%%%%%%%%%%%%%%%%%%%%%%

\noindent{\bf 3. Poincar\'e Polynomials}

Chiral-chiral (chiral-antichiral) primary states of the full $N=2$ SCA
are states in the Neveu-Schwarz (NS) sector defined
by the condition $\D={Q\over 2}$, $\bar\D={\bar Q\over 2}$ ($\D={Q\over 2}$,
$\bar\D=-{\bar Q\over 2}$). In our conventions, barred
quantities refer to the right $N=2$ algebra.
For $CP_m$ coset models,
states satisfying $\D=Q/2$ are those of the form $^{\cite{gepnercmp}}$
\beq\label{chiral}
|\L,\L,(0)\ket =
|\sum_{i=1}^{m} n_i \w_i, \sum_{i=1}^{m} n_i \w_i,(0)\ket
\eeq
and their $\s$-transformed ones. For states like
(\ref{chiral}) the integers $N,M$ in equations (\ref{dim}), (\ref{carga}) are
zero $^{\cite{gepnercmp}}$, whereas for the $\s$-transformed ones they adjust
to give the same value of $\D$ and $Q$.

The set of chiral primary fields of an $N=2$ superconformal algebra
is known to possess a bigraded commutative algebra structure
$^{\cite{gepnercmp,refpoin}}$,
with the bigrading given by the left and right $U(1)$ supersymmetry charges
$Q$ and $\bar Q$. Associated to this (finite dimensional) chiral algebra,
the Poincar\'e polynomial can be defined as
\beq
\P(t,\bar t)=\sum_{chiral~states} t^{DQ} \;\bar t\,^{D\bar{Q}}
\eeq
Here $D$ is the smallest integer such that $DQ$ is an integer for all
states.

In this article we have computed Poincar\'e polynomials for those
$CP_m$ cosets with nondiagonal invariants which are relevant to
string compactification. All chiral primary states were constructed
with a computer program, and their $U(1)$ charge was thus
computed. Field identifications (\ref{fi}) had to be taken properly into
account.

If either $\cal{N}$ or $\cal{M}$ are diagonal,
then $Q=\bar Q$ and the polynomials
take the simple form
\beq
\P(t \bar t) = \sum_{|\L,\L,(0)\ket} {\N}_{\L,\L} {\M}_{\L,\L}
\left( t \bar t\right) ^{Dq(|\L,\L,(0)\ket)\over k+m+1}
\eeq
For example, for the coset $SU(3)_9/SU(2)_{10}\times U(1)\equiv (2,9)CA$
\footnotemark[1] \footnotetext[1]
{We denote the coset $\G_{k}/\SUM_{k+1}\times U(1)$
with invariants $\N$ and $\M$ by $(m,k)\N\M$.}
with invariants ${\N}_{\L,\bar\L}$= C(2,9) and
${\M}_{\l,\bar\l}=\delta_{\l,\bar\l}$=A(1,10)
we obtain the polynomial
\beq
\P(t \bar t)= 1+ 3(t\bar t)^2 + 4 (t\bar t)^3 + 3 (t\bar t)^4 + (t\bar t)^6
\hspace{1.2cm} D=4
\eeq

Also, if both $\N$ and $\M$ are diagonal we recover the result of
Ref.~\cite{butu}. For series invariants the polynomials can be cast in
compact expressions for arbitrary values of $k$. Examples for $CP_1$ and $CP_2$
are given in Table~II.
States with $Q\neq \bar Q$ only appear when both $\N$ and $\M$
are nondiagonal.  These models do not admit a Landau- Ginzburg description.
The Poincar\'e polynomials for some of these cases are listed in
Table~III $^{\cite{inftec}}$.

It can be seen that, although we are considering nondiagonal invariants for
$\SUM$, the field $C_{max}$ with charge $Q_{max}=\bar Q_{max}=c/3$ is always
in the theory and therefore the Poincar\'e polynomial possesses the duality
property $^{\cite{lvw,gepnercmp}}$
\beq
P({1\over t},{1\over {\bar t}})=(t \bar t)^{-cD/3} \,P(t,\bar t)
\eeq

Equivalences among some of the models with diagonal $\M$ have already been
listed in the literature $^{\cite{models,nos}}$. We further reconfirm them and
add the following identities (denoted by $\equiv$) among the Poincar\'e
polynomials corresponding to the models:
$$
\begin{array}{l}
(3,8)DD  \equiv (3,8)DC \equiv (3,8)DE \equiv (3,8)CD \equiv
(3,8)CE  \equiv (3,8)ED \equiv \nn\\

(3,8)EC  \equiv (3,8)EE \equiv (4,5)DD_2 \equiv (4,5)DC \equiv
(4,5)CD_2 \equiv (4,5)EC \nn\\
\end{array}
$$
$$
\begin{array}{ll}
(2,5)CF \equiv  (2,5)CA                &~~~~~ (2,9)DE  \equiv (2,9)EE  \nn\\

(2,9)CA \equiv  (2,9)CF \equiv (3,4)CA &~~~~~ (2,9)CE  \equiv (3,4)CC  \nn\\

(2,15)AE \equiv (3,5)AA                &~~~~~ (2,15)DE \equiv (3,5)AD \nn\\

(2,21)CA \equiv (2,21)CF               &~~~~~ (2,27)DE \equiv (3,6)D_2A\nn\\

(3,8)D_2E \equiv (4,5)DA               &~~~~~ (3,8)CC  \equiv (4,5)CC \nn\\

(m,k)AX  \equiv (m-1, k+1)XA \,\,(1,{k-m+1\over m})A  \nn
\end{array}
$$
where $X$ denotes an arbitrary invariant.

 ~~

%%%%%%%%%%%%%%%%%%%%%%%%%%%%%%%%%%%
%\section{4D String Construction}  %
%%%%%%%%%%%%%%%%%%%%%%%%%%%%%%%%%%%

\noindent{\bf 4. 4D String Construction}

To make a string theory out of these models the Gepner
construction has to be followed: each sector (left and right moving)
will be a product of spacetime bosons and fermions times the
product of internal $N=2$ coset fields, such that $c_{int}=9$.
To obtain $N=1$ spacetime supersymmetry, a projection over
states with odd integer $U(1)$ charge $Q$ must be performed.

By building the internal sector as a tensor product of $r$ $CP_m$ coset
models$^{\cite{ks}}$, states in the complete internal theory are given by
\beq
\otimes_{i=1}^r  |\L_i,\ls_i,q_i,\Ls_i\ket
\eeq
It is useful to denote each state in the full theory by a vector
\beq
V=(\Ls_0;q_1,\dots,q_r;\Ls_1,\dots,\Ls_r)
\eeq
where $\Ls_0$ is an $SO(2)$ weight.
We also introduce the scalar product
\beq
V\cdot V'= \sum_{i=0}^r \Ls_i \Ls'_i
-\sum_{i=1}^r {q_i q'_i\over 2 \eta_i (k_i+m_i+1)}
\eeq
and the vectors
\beqa
\beta_0&=&(\bar s; \eta_1,\dots,\eta_r;s_1,\dots,s_r)\nn\\
\beta_i&=&(v;0,\dots,0;0,\dots,v,\dots,0) \hspace{1.5cm}
\mbox{($v$ in the $i$-th. position)}
\eeqa
where $\eta_i= {1\over 2} m_i(m_i+1)$.

The supersymmetry projection and aligned boundary conditions (R-R and
NS-NS) are accomplished by
\beqa
Q(\bar V)= 2\beta_0\cdot \bar V= \mbox{odd integer}\label{qentero}\\
2\beta_i\cdot \bar V= \mbox{even integer}
\eeqa
For Neveu-Schwarz states condition (\ref{qentero}) amounts to integer
internal
charge $Q_{int}$. To maintain modular invariance, twisted sectors
must be included. This is achieved by the condition
\beq \label{vect}
V=\bar V+s\beta_0+\sum_{i=1}^r n_i\beta_i
\eeq
with $s$ and $n_i$ integers. It is understood that
a given state is allowed whenever both invariants $\N$ and $\M$
are nonvanishing and its multiplicity is given by the product of
the modular coefficients.
Equation (\ref{vect}) means that for NS-NS states,
in the $s$-th twisted sector the condition $q_i=\bar q_i$
is replaced by
\beq
q_i-\bar q_i=2s\eta_i
\label{qtwist}
\eeq
However, since the shift vector $\beta_0$ has an even integer charge,
$Q(\beta_0)=2$,
the relation $Q_{int}-\bar Q_{int} \in Z$ still holds for $c_{int}\in 3Z$.

The heterotic construction is implemented by replacing left spacetime
fermions by internal free bosons with central charge
$c=24$ to cancel the bosonic anomaly. Modular invariance of the
theory follows from the isomorphism between representations of $SU(2)$
and the new gauge group $E_8\times SO(10)$ under the modular group.
With the supersymmetry projection the gauge group gets enlarged
 from $E_8\times SO(10)$ to $E_8\times E_6$.

Massless matter (antimatter) fields belong to the ${\bf 27}$ (${\bf\ol{27}}$)
representation of $E_6$, which decomposes as
${\bf 27}={\bf 10}_{\bf 1}+{\bf 16}_{\bf -1/2}+{\bf 1}_{\bf -2}$
(${\bf \ol{27}}={\bf 10}_{\bf -1}+{\bf \ol{16}}_{\bf 1/2}+{\bf 1}_{\bf 2}$)
of $SO(10)\times U(1)$.
The number of chiral generations $N_{27}$ ($N_{\ol{27}}$) may be obtained by
looking at the coupling of a left $SO(2)$ scalar with a right $SO(10)$ vector.
The masslessness condition implies that the $27$ correspond to chiral states
with $\D_{int}=Q_{int}/2=1/2$, $\ol \D_{int}=\ol Q_{int}/2=1/2$
and the $\ol{27}$
correspond to antichiral states with
$\D_{int}=Q_{int}/2=1/2$, $\ol \D_{int}=-\ol Q_{int}/2=1/2$. Alternatively,
an equivalent counting of $\ol{27}$  can be achieved by considering the right
$SO(10)$ scalar, that is, states with $\ol \D_{int}=\ol Q_{int}/2=1$.
The net number of generations is then given by $N_{gen}=|N_{27}-N_{\ol{27}}|$.

If the internal theory corresponds to a compactification on a Calabi-Yau
manifold, the $\ol{27}$ and ${27}$ are related to $b_{1,1}$ and $b_{1,2}$
respectively, where $b_{p,q}$ are the number of harmonic $(p,q)$ forms
on the manifold. The Euler characteristic  is, for $c_{int}=9$,
\beq\label{euler}
\chi= \sum_{p,q=0}^3 (-1)^{p+q} b_{p,q} = 2 (b_{1,1}-b_{1,2})
\eeq

A way of computing $\chi$ relies on the knowledge of the previously
constructed Poincar\'e polynomials. Each superstring model, with its internal
sector composed of a tensor product of $CP_m$ cosets, will thus be
characterized by a Poincar\'e polynomial
\beq
\P(t,\bar t)=\prod_{i=1}^r \P_i(t^{D/D_i},\bar t\;^{D/D_i})
\eeq
where $\P_i$ are the Poincar\'e polynomials for each coset model,
$D_i$ are the least integers (in each model) such that $D_iQ_{int}$ is an
integer and $D$ is the least common multiple of all $D_i$.

However, this polynomial only counts states in the untwisted sector. To include
twisted sectors we may also define ``twisted polynomials", $\P^{(s)}(t,\bar t)=
\prod_{i=1}^r \P_i^{(s)}(t^{D/D_i}, \bar t^{D/D_i})$. These are
constructed by summing over left and right chiral states coupled according to
the modular invariants and satisfying condition (\ref{qtwist}),
\beq
\P_i^{(s)}=\sum_
{q_i-\bar q_i=2s\eta_i}
t^{DQ_i}\,\bar t\;^{D\bar Q_i}
\eeq

The number of $27$ and $\ol{27}$ are obtained as the coefficients of
$t^D \bar t^D$ and $t^D\bar t^{2D}$ in
\beq
\P^{sum}=\sum_{s=0}^{D-1} \P^{(s)}
\eeq

If one is only interested in the Euler characteristic, these expressions can be
computed employing a formula derived by
Vafa$^{\cite{vafa}}$ for Landau-Ginzburg models
and extended to more general $N=2$ superconformal theories by
Buturovic$^{\cite{butu}}$,
\beq
\label{gener}
N_{gen}={1\over 2D}\sum_{r,s=0}^{D-1} P_{r,s}
\eeq
where
\beq
P_{r,s}= \left.Tr\left\{(-1)^{rc/3} e^{2i\pi r J_0} e^{i\pi(Q-\bar Q)}\right\}
\right|_{{R}^{(s)}_0}
\label{pol}\eeq
and ${R}^{(s)}_0$ is the Ramond ground state of the $s$-th twisted sector.
Summation over $s$ includes all twisted sectors and
summation over $r$ projects over integer $Q_{int}$. The factor
$e^{i\pi (Q_{int}-\bar Q_{int})}$ takes into account the term $(-1)^{p+q}$
of equation (\ref{euler}).

Due to modular invariance, it is possible ~$^{\cite{vafa,butu}}$ to write the
contribution of the twisted sectors in terms of the untwisted one
as $P_{r,s}=P_{x_{r,s},0}$ with
$x_{r,s}$ the greatest common divisor of $r,s$. By spectral flow, the Ramond
ground state $R_0$ can be related to the chiral ring ${\cal R}$
of the NS sector.
It thus follows that the trace (\ref{pol}) over the
twisted Ramond ground state is just the Poincar\'e polynomial of the
theory evaluated at particular values of the parameters $t{,}\bar t$:
\beq\label{tvalue}
P_{x,0}= \P(t=e^{2i\pi x/D+i\pi/D};\bar t=e^{-i\pi/D})
\eeq
Notice that since we are considering nondiagonal modular invariants
we keep the term $e^{i\pi(Q-\bar Q)}$ in the untwisted sector.

When considering moddings by discrete symmetries we will need expressions
relating different twisted sectors for each internal coset theory. We
therefore rewrite explicitly equation (\ref{gener}) in terms of twisted
polynomials,
\beq
N_{gen}={1\over 2D}\sum_{r,s=0}^{D-1} \prod_{i=1}^r
\P^{(s)}_i(t=e^{2i\pi r/D+i\pi/D},
\bar t=e^{-i\pi/D})
\eeq

 ~~

%%%%%%%%%%%%%%%%%%%%%%%%%%%%%%%%%%%%%%%%%%%%
%\section{Moddings by Discrete Symmetries}  %
%%%%%%%%%%%%%%%%%%%%%%%%%%%%%%%%%%%%%%%%%%%%

\noindent{\bf 5. Moddings by Discrete Symmetries}

As is well known$^{\cite{gepner,modd,gq}}$,
$CP_m$ cosets have a discrete $Z_{k+m+1}$ symmetry
which, when taken into account, may reduce the net number of
generations of the corresponding string compactification. Dividing by
$Z_N$ is equivalent to introducing twisted boundary conditions which can
be included in the corresponding Poincar\'e polynomials as we will
now explain.

These moddings are characterized by a vector
\beq
\Gamma=(0;\gamma_1\eta_1,\dots,\gamma_r\eta_r;0,\dots;0)
\eeq
where $\gamma_i$ are integers satisfying
\beq
2\beta_0\cdot \Gamma = - \sum_{i=1}^r {\gamma_i\eta_i\over k_i+m_i+1}
= integer
\eeq

States in the twisted sectors now verify
\beq
V=\bar V+s\beta_0+\sum_{i=1}^r n_i\beta_i+2x\Gamma
\eeq
and the generalized GSO projection over states with integer $Q_{int}$ is now
given by the conditions $^{\cite{modd}}$
\beqa
2\beta_0\cdot \bar V&=&odd\\
2\beta_i\cdot \bar V&=&even\\
-\Gamma(2\bar V+2x\Gamma)&=&integer\label{yproj}
\eeqa

A modular invariant partition function is constructed by summing over
all twisted sectors and implementing the above projections. Keeping in mind
that only states allowed by the invariants $\N$ and $\M$
(with its corresponding multiplicity) should be coupled and defining
\beqa
Z&&(s,n_i,x;r,m_i,y)= \\
&&=\sum_V (-1)^{s+r}
e^{-2i\pi V(r\beta_0+\sum_i m_i\beta_i)}\,
e^{-2i\pi y\Gamma(2V+2x\Gamma)}\,
\chi_V\,    \chi^*_{V+s\beta_0+\sum_i n_i\beta_i+2x\Gamma}\nn
\eeqa
the partition function is given by
\beq
Z={1\over D^{r+1}M}
\sum_{r,s,n_i,m_i}^{D-1} \sum_{x,y=0}^{M-1} Z(s,n_i,x;r,m_i,y)
\eeq

The summation over $y$ implements condition (\ref{yproj}) and summation over
$x$ is such that all sectors twisted by $\Gamma$ are included. Therefore,
both sums should range from $0$ to $M-1$, $M$ being the least integer such
that $M\gamma_i=I$ mod$(k_i+m_i+1)$ for all~$i$, with $I$ a positive
integer.

The computation of the Euler characteristic for the associated compactified
manifolds can now be carried out with the knowledge of the twisted Poincar\'e
polynomials.
To generalize eq. (\ref{pol}) by including the moddings,
we introduce the quantity
\beq\label{polmod}
P_{r,y;s,x}=
Tr\left\{(-1)^{rc/3} e^{i\pi(Q-\bar Q)}e^{2i\pi r J_0}
e^{2i\pi y\sum_i \gamma_i(q_i+\bar q_i)/2(k_i+m_i+1)}
\right\}_{R^{(s,x)}_0}
\eeq
$R^{(s,x)}_0$ refers to the Ramond ground state such that
\beq
V-\bar V= 2s\beta_0+2x\Gamma
\eeq

Equation (\ref{polmod}) can be factorized as
\beqa\label{polfact}
P_{r,y;s,x}&=&
\prod_i {(P_i)}_{r,y;s,x}=\nn\\
&=&Tr\left\{\prod_i\left[ (-1)^{rc_i/3}e^{i\pi(Q_i-\bar Q_i)}e^{2i\pi r
{J_0}_i}
e^{2i\pi y\gamma_i(q_i+\bar q_i)/2(k_i+m_i+1)}
\right]_{{R_0}^{(s,x)}_i}\right\}
\eeqa
where ${R_0}_i^{(s,x)}$ consists of the states satisfying the condition
\beq \label{twistmod}
(q_i-\bar q_i)=2 \eta_i(s+x\gamma_i)
\eeq
This condition allows us to replace $q_i+\bar q_i$ by
$2q_i-2\eta_i(s+x\gamma_i)$ in equation (\ref{polfact}). Consequently,
%With this replacement and as a consequence
%of the fact that $\sigma(q_i)=q_i+k_i+m_i+1$,
it is also possible to replace
$q_i/(k_i+m_i+1)$ by $Q_i$. Furthermore,
${(P_i)}_{r,y;s,x}={(P_i)}_{r,y;s+x\gamma_i,0}$.

The number of chiral generations is then given by
\beq
N_{gen}= {1\over 2MD}\sum_{r,s=0}^{D-1}\sum_{x,y=0}^{M-1} P_{r,y;s,x}
\eeq
Again, $P_{r,y;s,x}$ may be written in terms of products of
twisted Poincar\'e polynomials as
\beqa
P_{r,y;s,x}&=&\prod_{i=1}^r {(P_i)}_{r,y;s,x}=  \nn\\
&=&\prod_{i=1}^r e^{-2i\pi y\gamma_i^2 x \eta_i/(k_i+m_i+1)}\,
\P_i^{(s+x\gamma_i)}(t=e^{2i\pi r/D+i\pi/D+2i\pi y\gamma_i/D};
\bar t=e^{-i\pi/D})
\eeqa

 ~~

%%%%%%%%%%%%%%%%%%%%%%%%%%%%%%%%%%%%
%\section{Results and Conclusions}  %
%%%%%%%%%%%%%%%%%%%%%%%%%%%%%%%%%%%%

\noindent{\bf 6. Results and Conclusions}

In section 3 we have constructed Poincar\'e polynomials for $CP_m$ coset models
with a large class of nondiagonal modular invariants. The polynomials are
concise expressions containing information about the elements of the chiral
algebra of the model.  Equivalences among models can be more easily
established by observing whether their Poincar\'e polynomials coincide.
When constructing superstring models in four dimensions as discussed in section
4, the massless spectrum may be obtained from the Poincar\'e polynomial,
thus bypassing the explicit construction of states  case by case.

We have used the formulas derived above to compute the number of generations
for string compactifications based on tensor products of $CP_m$ coset
models, containing at least one coset with nondiagonal $\SUM$ invariant. A full
list of results is given in \cite{inftec}. These results do not differ
much from those previously obtained in the
literature~$^{\cite{models,schell,nos,modd}}$ for less
general models. The number of generations ranges from zero to
360 for the 1144 models considered.
$N_{gen}$ either zero or multiple of 8, 12, 18 are
the numbers more often found. Models with $0<N_{gen}<12$ are listed
in table~IV.

It is interesting to notice that some coset models possess Poincar\'e
polynomials that vanish when evaluated at the appropriate values of
$t,\t$ given by equation~(\ref{tvalue}). Therefore, all string
compactifications
containing these cosets as internal theories will have vanishing number of
generations. They may correspond to models with $N_{27}=N_{\ol{27}}=5, 9, 13$
as well as to compactifications on $K_3\times T$ with
$N_{27}=N_{\ol{27}}=21$. Models with vanishing polynomials are displayed
in table~V.

In order to search for superstring models with lower number of generations
we have also applied the results of section~5 to the aforementioned coset
models. Taking into account single moddings by phase symmetries we studied
4819 models. We
found that
more than half of these have zero generations. We also
encountered 2, 18, 59 and 367 models with $N_{gen}$=4, 6, 8 and 12,
respectively.

 ~~

%%%%%%%%%%%%%%%%%%%%%%%%%%%%%%
% \section{Acknowledgments}  %
%%%%%%%%%%%%%%%%%%%%%%%%%%%%%%

\noindent{\bf Acknowledgments}

We thank A. Font for her contribution during the initial stage of this
work. Comments by Prof. D. Gepner are also very gratefully
acknowledged. Two of us (G.A. and I.A.) thank IAFE and C.N. thanks CAB
for hospitality.

%%%%%%%%%%%%%%%%%%%%%%%%%%%%%%%%%%%%%%%%%%%%%%%%%%%%%%%%%%%%%%%%%%%%%%%%%%%%

\begin{table}
\caption{\noindent
Modular invariant for $SU(6)$ at level 8 from the conformal embedding
into $SO(21)$ at level 1. $^a$}
\begin{tabular}{c}
$\begin{array}{rl}
C(5,8)=|\c{00000}+\c{20002}+&\c{21012}+\c{03030}
+\c{03103} + \\
&\c{30130}+\c{30203}+\c{12221}+\c{02420}+\c{00800}|^2  +  \\
|\c{00024}+\c{01301}+\c{02030}+\c{10122}+&\c{00008}+\nn\\
&\c{31031}+\c{22210}+\c{30302}+\c{24200}+\c{08000}|^2  +
\bc + \nn\\
|\c{00002}+\c{20012}+\c{21030}+\c{21103}+&\c{03121}+
\c{30221}+\c{12320}+\c{02600}+\c{04004}|^2  + \bc  + \\
|\c{00012}+\c{20030}+\c{20103}+\c{22004}+&\c{30320}+
\c{03220}+\c{04022}+\c{12500}+\c{21121}|^2  +  \bc  + \\
|\c{40012}+\c{30100}+\c{12102}+\c{03000}+&\c{12140}+
\c{02212}+\c{00430}+\c{00503}+\c{22022}|^2  +  \bc  + \\
|\c{00121}+\c{01004}+\c{21022}+\c{30005}+&\c{21400}+
\c{22121}+\c{04301}+\c{05030}+\c{20220}|^2  +  \bc  + \\
|\c{00220}+\c{01022}+\c{10005}+\c{20400}+&\c{30023}+
\c{22301}+\c{23030}+\c{05210}+\c{21121}|^2  +  \bc  + \\
|\c{60000}+\c{32001}+\c{22103}+\c{12110}+&\c{10312}+
\c{03022}+\c{01232}+\c{00260}+\c{00400}|^2  +  \bc  + \\
|\c{00123}+\c{02210}+\c{10302}+\c{11031}+&\c{00026}+
\c{23200}+\c{31211}+\c{26000}+\c{40040}|^2  +  \bc  + \\
|\c{40220}+\c{32201}+\c{25000}+\c{20040}+&\c{03200}+
\c{01032}+\c{00303}+\c{00125}+\c{11211}|^2  +  \bc  + \\
|\c{00050}+\c{21210}+\c{14001}+\c{10221}+&\c{43010}+
\c{10043}+\c{01214}+\c{50300}+\c{02202}|^2  +  \bc
\end{array}$
\end{tabular}
\tablenotes{$^a$ $|cr|$ denotes the sum of conjugate representations}
\end{table}
%%%%%%%%%%%%%%%%%%%%%%%%%%%%%%%%%%%%%%%%%%%%%%%%%%%%%%%%%%%%%%%%%%%%%%%%%%%%%
\begin{table}
\caption{Poincar\'e polynomials for $CP_1$ and $CP_2$ models with series
invariants}
\begin{tabular}{llr}
Model & Poincar\'e polynomial&~\\
\tableline
$(1,k)D,F$ & $P(t\t)= \sum_{i=0}^{k/2} (t\t)^{2i} + (t\t)^{k/2}$& $k$ even\\
\tableline

$(2,k)DA$& $P(t\t)=\sum_{n=0}^{{k\over 3}-1}(\left[{3n\over 2}\right]+1)
[(t\t)^{3n}+(t\t)^{2k-3n}]+(\left[{k\over 2}\right]+3)(t\t)^k$& $k\in 3N$$^b$\\
\tableline

$(2,k)AD$ & $P(t\t)=\sum_{n=0}^{{k-1\over 2}}(n+1)
[(t\t)^{n}+(t\t)^{k-n}]+\sum_{n={k+1\over 4}}^{{3k-1\over 4}}
(t\t)^{n}$ & $k=4j-1, j\in Z$\\
\tableline

$(2,k)AF$ & $P(t\t)=\sum_{n=0}^{{k-1\over 2}}(n+1)
[(t\t)^{2n}+(t\t)^{2k-2n}]+\sum_{n={k-1\over 4}}^{{3k+1\over 4}}
(t\t)^{2n+1}$ & $k=4j+1; j\in Z$\\
\tableline

$(2,k)DD$ & $P(t,\t)=\sum_{n=0}^{{k-3\over 6}}(3n+1)
[(t\t)^{n}+(t\t)^{k/3-n}]+$&$k=4j-1=3j';$\\
 ~&~~~~~~~$
\sum_{n={k+9\over 12}}^{{3k-9\over 12}}
(t\t)^{n}+t^{{k-3\over 12}}\t^{{3k+3\over 12}}
+t^{{3k+3\over 12}}\t^{{k-3\over 12}}
$ &  $j,j'\in Z$\\
\tableline

$(2,k)DF$ & $P(t,\t)=\sum_{n=0}^{{k-3\over 6}}(3n+1)
[(t\t)^{2n}+(t\t)^{2k/3-2n}]+$&$k=4j+1=3j';$\\
 ~&~~~~~~~$
\sum_{n={k+3\over 12}}^{{3k-15\over 12}}
[(t\t)^{2n+1}] +t^{{k-3\over 6}}\t^{{3k+3\over 6}}
+t^{{3k+3\over 6}}\t^{{k-3\over 6}}
$ &  $j,j'\in Z$\\

\end{tabular}
\tablenotes{$^b$ $[.]$ denotes the integer part.}
\end{table}
%%%%%%%%%%%%%%%%%%%%%%%%%%%%%%%%%%%%%%%%%%%%%%%%%%%%%%%%%%%%%%%%%%%%%%%%%%%%%%
\begin{table}
\caption{Poincar\'e polynomials  for  coset  models with exceptional $\N$
and $\M$}
\begin{tabular}{ll}
Model & Poincar\'e polynomial \\
\tableline
$(2,9)CE, ~(3,4)CC$ &
$\P(t,\t)=
1+t^2+t\bar t+t\bar t^3+\bar t^2+(t\bar t)^2+t^3\bar t+(t\bar t)^3$\\

$(2,9)EE$ &
$\P(t,\t)=1+t\bar t+5(t\bar t)^2+4(t\bar t)^3+5(t\bar t)^4+(t\bar t)^5+
(t\bar  t)^6$\\

$(3,8)CC, ~(4,5)CC$ &
$\P(t,\t)=1+2t+t^2+2\bar t+
4t\bar t+2t^2\bar t+\bar t^2+2t\bar t^2+(t\bar t)^2$\\

$(3,8)CE, ~EC, ~EE,$ &~ \\
 ~~~~~~~~~$(4,5)EC$ &
$\P(t,\t)= 1+t^2+20 t\bar t+\bar t^2+(t\bar t)^2$ \\

$(4,7)CC$ &
${\cal{P}}(t,\bar t)=1+2(t\bar t)^2+(t\bar t)^3+(t\bar t)^4+2(t\bar t)^5
+(t\bar t)^7+$\\
 ~&
 ~~~~~~$+t^3+\bar t^3+2t^2\bar t^5+2t^5\bar t^2
+t^4\bar t^7+t^7\bar t^4
$\\

$(4,7)CE$ &
${\cal{P}}(t,\bar t)=1+4(t\bar t)^2+t^2\bar t^5+8(t\bar t)^3+8(t\bar t)^4
+t^5\bar t^2+4(t\bar t)^5+(t\bar  t)^7$\\

$(5,6)CC$ &
$\P(t,\t)=1+2t^2+t^4+t\bar t+2 t\bar t^3+t\bar t^5+2\bar t^2+4(t\bar t)^2
+2 t^2\bar t^4+$\\
 ~&
 ~~~~~~$+2 t^3\bar t+4(t\bar t)^3+2 t^3\bar t^5+\bar  t^4+2 t^4\bar t^2
+(t\bar t)^4+t^5\bar t+2 t^5\bar t^3+(t\bar t)^5$\\

$(5,6)\tilde CC$ &
$\P(t,\t)=1+t^2+t\bar t+t\bar t^3+\bar t^2+2 (t\bar t)^2+t^2\bar t^4
+t^3\bar t+$\\
 ~&
 ~~~~~~$+2 (t\bar t)^3+t^3\bar t^5+t^4\bar t^2+(t\bar t)^4+t^5\bar t^3+
(t\bar t)^5$\\

$(6,7)CC$ &
$\P(t,\t)=1+3 t+3 t^2+t^3+3\bar t+9  t\bar t+9 t\bar t^2+3 t\bar t^3+
3\bar t^2+9 t^2\bar t+$\\
 ~&
 ~~~~~~$+9(t\bar t)^2+3 t^2\bar t^3+\bar t^3+3 t^3\bar t+
3 t^3\bar t^2+(t\bar t)^3$\\
\end{tabular}
\end{table}
%%%%%%%%%%%%%%%%%%%%%%%%%%%%%%%%%%%%%%%%%%%%%%%%%%%%%%%%%%%%%%%%%%%%%%%%%%%%%%
\begin{table}
\caption{Nondiagonal $CP_m$ coset models with $0<N_{gen}<12$.}
\begin{tabular}{lrrr}
Model~~~~~~~~~~~~~~~~~~~~~~~~~~~~~~~~~~~~~~~~~~~~~~~~~~~~~~~~~~~~~~~~~~~~~~~~
& $N_{27}$ & $N_{\ol{27}}$ &  $N_{gen}$\\
\tableline
(2,9)EF (1,3)A (1,18)A       &   25    &   21    &   4\\
(2,9)DF (2,9)AA              &   28    &   22    &   6\\
(2,9)DA (2,9)AF              &   28    &   22    &   6\\
(2,9)DF (2,9)DA              &   28    &   22    &   6\\
(2,9)DF (1,4)A (1,10)A       &   28    &   22    &   6\\
(3,4)FC (1,6)A (1,6)A        &   29    &   21    &   8\\
(3,4)FC (1,6)F (1,6)F        &   29    &   21    &   8\\
(3,4)$D_2$C (3,4)AC          &   27    &   19    &   8\\
(3,4)FC (3,4)AC              &   19    &   27    &   8\\
(3,4)$D_2$C (3,4)$D_2$C      &   27    &   19    &   8\\
(3,4)FC (3,4)$D_2$C          &   19    &   27    &   8\\
(3,4)FC (3,4)FC              &   27    &   19    &   8\\
(3,4)AC (2,9)EF              &   24    &   16    &   8\\
(3,4)$D_2$C (2,9)EF          &   24    &   16    &   8\\
(3,4)FC (2,9)EA              &   24    &   16    &   8\\
(6,7)AC                      &   16    &   8     &   8\\
(6,7)DC                      &   16    &   8     &   8
\end{tabular}
\end{table}
%%%%%%%%%%%%%%%%%%%%%%%%%%%%%%%%%%%%%%%%%%%%%%%%%%%%%%%%%%%%%%%%%%%%%%%%%%%%%%
\begin{table}
\caption{Models with Poincar\'e polynomials vanishing at
$t=e^{2i\pi x/D+i\pi/D};\bar t=e^{-i\pi/D}$}
\begin{tabular}{ll}
Model & Poincar\'e polynomial \\
\tableline

(2,3)DD & $\P(t,\t)= (1+t) (1+\t) $ \\
(2,9)CE, (3,4)CC & $\P(t,\t)= (1+t^2) (1+\t^2) (1+t \t) $\\
%(3,4)CC & $\P(t,\t)= (1+t^2) (1+\t^2) (1+t \t) $\\
(3,8)CC, (4,5)CC& $\P(t,\t)= (1+t)^2 (1+\t)^2 $\\
(5,6)CC & $\P(t,\t)= (1+t^2)^2 (1+\t^2)^2 (1+t \t) $\\
(5,6)$\tilde{C}$C & $\P(t,\t)= (1+t^2) (1+\t^2) (1+t \t) (1+t^2 \t^2) $\\
(6,7)CC & $\P(t,\t)= (1+t)^3 (1+\t)^3$
\end{tabular}
\end{table}
%%%%%%%%%%%%%%%%%%%%%%%%%%%%%%%%%%%%%%%%%%%%%%%%%%%%%%%%%%%%%%%%%%%%%%%%%%%%
%%%%%%%%%%%%%%%%%%%%%%%%%%%%%%%%%%%%%%%%%%%%%%%%%%%%%%%%%%%%%%%%%%%%%%%%%%%%

%%%%%%%%%%%%%%%%%%%%%%%%%%%
% \section{REFERENCIAS}   %
%%%%%%%%%%%%%%%%%%%%%%%%%%%

\end{document}